\documentclass[aip,jap,superscriptaddress,amsmath,preprintnumbers,amssymb,floatfix,showpacs,showkeys,12pt,preprint]{revtex4-1}
\usepackage{graphicx}
\usepackage{dcolumn}
\usepackage{bm}
\usepackage{subfig}
\def\dd{\hbox{\rm d}}

\def\N{N(\alpha,\beta)}
\def\Q{Q(t,\alpha_0,\alpha,\beta_0,\beta)}
\def\U{U(t,\alpha,\beta)}
\def\S{S(t,\alpha_0,\alpha,\beta_0,\beta)}
\begin{document}
\title{Hot-carrier trap-limited transport in switching chalcogenides}
\author{Enrico Piccinini}
\email{enrico.piccinini@unimore.it}
\affiliation{"E. De Castro" Advanced Research  Center on Electronic Systems (ARCES), Universit\`a di Bologna, Via  Toffano 2/2, I-40125 Bologna, Italy}
\author{Andrea Cappelli}
\affiliation{Dipartimento di Scienze Fisiche, Informatiche e Matematiche, Universit\`a di Modena e Reggio Emilia, Via Campi 213/A, I-41125 Modena, Italy}
\author{Fabrizio Buscemi}
\affiliation{"E. De Castro" Advanced Research Center on Electronic Systems (ARCES), Universit\`a di  Bologna, Via  Toffano 2/2, I-40125 Bologna, Italy}
\author{Rossella Brunetti}
\affiliation{Dipartimento di Scienze Fisiche, Informatiche e Matematiche, Universit\`a di Modena e Reggio Emilia, Via Campi 213/A, I-41125 Modena, Italy}
\author{Daniele Ielmini}
\affiliation{Dipartimento di Elettronica e Informazione, Politecnico di Milano
, Piazza Leonardo da Vinci 32, I-20133 Milano, Italy}
\author{Massimo Rudan}
\affiliation{"E. De Castro" Advanced Research Center on Electronic Systems (ARCES), Universit\`a di  Bologna, Via  Toffano 2/2, I-40125 Bologna, Italy}
\author{Carlo Jacoboni}
\affiliation{Dipartimento di Scienze Fisiche, Informatiche e Matematiche, Universit\`a di Modena e Reggio Emilia, Via Campi 213/A, I-41125 Modena, Italy}
\begin{abstract}
Chalcogenide materials have received great attention in the last decade owing to their  application in new memory systems. Recently, phase-change memories have, in fact, reached the early stages of production. In spite of the industrial exploitation of such materials, the physical processes governing the switching mechanism are still debated.
In this paper we work out a complete and consistent model for transport in amorphous chalcogenide materials based on trap-limited conduction accompanied by carrier heating. A previous model is here extended to include position-dependent carrier concentration and field, consistently linked by the Poisson equation. 
The results of the new model reproduce the experimental electrical characteristics and their dependences on the device length and temperature. Furthermore, the model provides a sound physical interpretation of the switching phenomenon and is able to give an estimate of the threshold condition in terms of the material parameters, a piece of information of great technological interest.
\end{abstract}
%
%
\maketitle
\newpage
\section{Introduction}

With the recent introduction of the 22 nm node in the fabrication process, the semiconductor industry seems to be very close to its technological limit. According to the International Technology Roadmap for Semiconductors,\cite{ITRS} it could eventually be possible to scale down the actual devices to the next 1x generation only by redesigning either the device, or the productive process, or both. It will be more and more complicated, if not impossible, to continue this trend much further.\cite{Zhirnov,ITRS} Alternatively, new materials have to be explored and different architectural solutions implemented. 

In the memory technology, non-silicon materials are being widely investigated in order to introduce faster, more scalable, and reliable devices. The present frontier of the research is represented by metal-oxide resistive RAMs and conductive-bridge RAMs,\cite{ITRS} while Phase-Change Memory (PCM) prototypes have been studied in the last years\cite{Kau} and are now in the early production stage.\cite{Derhacobian,Wong} Phase-change materials like chalcogenides are known from the early 1960s\cite{Ovshinsky} and show the property of an easy, reversible transition between crystalline and amorphous phases, characterized by significant changes in optical reflectivity and electrical resistivity.\cite{Wuttig} Due to high optical contrast between the two phases, these materials have been employed for optical storage since the mid 1990s.\cite{Yamada} In the new century, the strong difference in resistivity characterizing the two phases pushed chalcogenide materials as suitable candidates for solid-state nonvolatile memories. Furthermore, some chalcogenide glasses also feature an ovonic threshold-switching in the amorphous phase, which implies a negative differential resistance (NDR) in the current-voltage characteristic before the phase change takes place. Even though the first PCM arrays have already been released to the market, the knowledge of the physical process governing the ovonic switching mechanism is still a step behind. The availability of a theory able to identify the threshold point and predict the device behavior under given operating conditions is still sought by scientists and engineers to tailor the materials and the device set up. 
 
The first microscopic interpretation of the switching behavior was due to Adler and coworkers,\cite{AdlerRMP,AdlerJAP} who supposed the creation of a micrometer-wide low-resistance filament in the amorphous matrix, thus reducing the resistance of the device. Later on, the switching behavior was also found for sub-micrometer devices,\cite{Lai} and this gave rise to alternative interpretations. Using the standard macroscopic quantities of the theory of transport in semiconductors, like concentrations, velocities, and mobilities, it is possible to explain the switching in terms of drift-diffusion with impact ionization,\cite{Pirovano} or cooperative detrapping.\cite{Rudan}

Alternative interpretations are due to Karpov and coworkers,\cite{Karpov,Simon} and to Ielmini and coworkers.\cite{IelminiJAP,IelminiPRB}  The interpretation provided by Karpov and coworkers preserves the idea of conductive filaments as responsible for the switching behavior. From an energetic balance, they derive a model based on nucleation and growth of a thin crystalline filament that progressively expands in the amorphous matrix until it connects the electrodes, thus dramatically reducing the electrical resistance of the device. 

On the other hand, Ielmini and coworkers proposed a thermally-assisted trap-limited conduction mechanism, where the switching is ascribed to the increase in the average kinetic energy of the carriers through the device, as a result of the balance between the field-induced energy gain and the energy relaxation due to the scattering with phonons. A non-uniformity of the electric field is also found. This model has the capability of interpreting not only the electrical characteristic of a memory cell, but also its dependence on thickness and temperature, which suggests a thermally-activated conduction mechanism. Similarly, it has been shown through Monte Carlo simulations that a switching behavior can be triggered by a space-charge accumulation near the contacts, if a field-enhanced hopping conduction is considered.\cite{Buscemi}  

According to Ref.\ \onlinecite{IelminiPRB} a non-uniformity in the electric field across the device is found near and above the switching condition, which must be sustained by a non-negligible positive charge in the region close to the cathode. However, the equations used in Ref.\ \onlinecite{IelminiPRB} do not include the effect of the variable concentration of the active carriers across the device, which is instead approximated as a constant. In the present paper we use the ideas presented in Ref.\ \onlinecite{IelminiPRB} as a starting point, and we work out a complete and consistent model which includes diffusion and achieves self-consistency between charge distribution and field. The results of this theoretical development reproduce the electrical characteristics and their dependences on the geometrical scaling factors and temperature, and provide a sound physical interpretation of the results.

\section{The model}\label{model}

The presence of defects inherent to an amorphous material implies the existence of a number of localized trap states in the band gap. If the conduction were due to a pure hopping process, carriers would tunnel among traps until they reach the collecting contact. In the so called \emph{trap-limited conduction} regime, instead, carriers undergo continuous trapping-detrapping processes by which they absorb and release energy in such a way that they overcome the potential barriers and move across the device. An intermediate mechanism (\emph{thermally-assisted tunneling}), where carriers absorb some energy and tunnel to the next trap, is also possible. The three mechanisms coexist, and the final transfer rate must encompass all of them.  Since pure hopping is mainly effective in the low-temperature range (i.e., well below room temperature),\cite{MottBook,ShklovskiiBook} we can neglect it as we are not interested in that temperature range. As for thermally-assisted tunneling, it was shown that its dependence upon temperature and field is similar to that of the trap-limited conduction process.\cite{IelminiPRB}  For these reasons, we develop the model for the trap-limited conduction case.

Let  $n_T$ be the trap concentration, assumed to be spatially uniform. Under equilibrium conditions, the concentration of electrons is independent of the position and given by the Fermi-Dirac statistics; the electron temperature $T$ coincides with the lattice temperature $T_0$. In the off-equilibrium conditions, we assume translational symmetry along the $x$ and $y$ directions, so that the quantities of interest depend only on $z$, the current direction. The carrier concentration $n(z)$ is still described by a Fermi distribution where a quasi-Fermi level $E_F(z)$ replaces the Fermi level $E_{F0}$ and $T(z)$ replaces $T_0$. 
A flat trap density of states $\Gamma=n_T/\Delta E_G$ is assumed inside the band gap $\Delta E_G=E_C(z)-E_V(z)$, where $E_C(z)$ and $E_V(z)$ denote the band gap edges. The band gap follows the potential profile along the $z$ axis. The number of carriers between $E_T$ and $E_T+\dd E_T$ is given by:
\begin{equation}\label{dnFermi}
\dd n=\frac{\Gamma}{1+\exp\left[\frac{E_T-E_F(z)}{kT(z)}\right]}\,\dd E_T,
\qquad
E_V(z)\leq E_T \leq E_C(z).
\end{equation}
The integration of Eq.\ (\ref{dnFermi}) over the band gap yields the carrier concentration at $z$:
\begin{equation}\label{nFermi}
n(z)=\int_{E_V(z)}^{E_C(z)}\frac{\Gamma}{1+\exp\left(\frac{E_T-E_F(z)}{kT(z)}\right)}\dd E_T=n_T-\Gamma kT(z) \ln\left(\frac{1+\exp\frac{E_C(z)-E_F(z)}{kT(z)}}{1+\exp\frac{E_V(z)-E_F(z)}{kT(z)}}\right).\end{equation}
The equilibrium value $n_0$ is obtained once $E_C(z)$, $E_V(z)$, $E_F(z)$ and $T(z)$ are replaced by their equilibrium values $E_{C0}$, $E_{V0}$, $E_{F0}$ and $T_0$:
\begin{equation}\label{neq}
n_0=n_T-\Gamma kT_0\ln\left(\frac{1+\exp\frac{E_{C0}-E_{F0}}{kT_0}}{1+\exp\frac{E_{V0}-E_{F0}}{kT_0}}\right).
\end{equation}
When $E_{F0}$ is sufficiently far from the band edges $E_{C0}$ and $E_{V0}$ and close to midgap, $n_0\approx n_T/2$. Under equilibrium conditions, the material is neutral. This situation may be obtained, for instance, by assuming donor-like (acceptor-like) traps and a negative (positive) compensating charge $n_0$.

The model is  in principle the same if one deals with electrons or holes. For this reason, in the followings we would rather term the two leads where carriers enter or leave the device as the \emph{injecting} contact, $z=0$, and the \emph{collecting} contact, $z=\ell$, respectively. For the sake of simplicity, we develop and discuss the model only for the case of electrons.

\begin{figure}
\includegraphics[width=.9\textwidth]{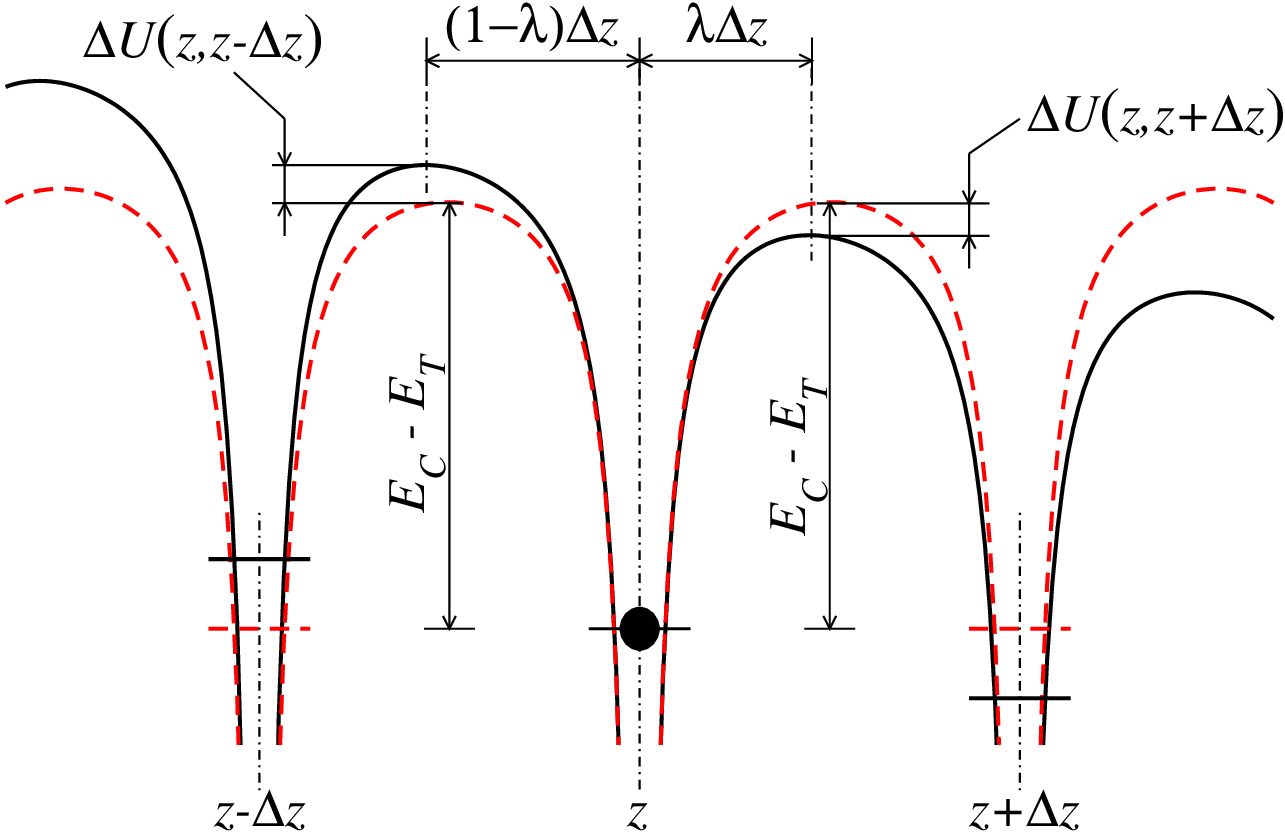}
\caption{(Color online) Schematic view of the energy profile for transitions from $z$ to $z+\Delta z$ and from $z$ to $z-\Delta z$. The dashed profile corresponds to the equilibrium condition, while the solid curve is obtained under the influence of a local field $F(z)<0$ that enhances the transitions towards larger $z$'s. Symbols are explained in the text.}
\label{schemapotenziale}
\end{figure}

Let $\Delta z$ be the average traveled distance between the sites of successive detrapping-trapping events. Following Ref.\ \onlinecite{IelminiJAP}, if the detrapping time is much longer than the traveling time, it can be taken equal to the transfer time $\tau$. For thermally-activated processes $\tau$ is exponentially dependent on the barrier height experienced by the carriers at the detrapping event. Thus, we can define two different times $\tau_{\rightarrow}$ and $\tau_{\leftarrow}$ that apply to the motion in the two directions, i.e., from $z$ to $z+\Delta z$ and from $z$ to $z-\Delta z$. The local field gives rise to opposite effects, as shown below:
\begin{equation}\label{tau}
\tau_{\rightleftharpoons}=\tau_0\exp\left[\frac{E_C(z)-E_T+\Delta U(z,z\pm\Delta z)}{kT_0}\right],
\end{equation}
where $\tau_0$ is a characteristic transfer time for the process at hand, and $\Delta U$ is the shift of the barrier height with respect to the equilibrium value due to the local electric field $F(z)$. With reference to Fig.\ \ref{schemapotenziale}, let $\lambda \Delta z$ be the distance of the maximum of the energy profile along the transition path from the trap having the lower $z$ coordinate, with $0<\lambda<1$. Using a first-order approximation, the shifts $\Delta U$ result:
\begin{equation}\label{DeltaU}
\Delta U(z,z+\Delta z)=qF(z) \lambda \Delta z, \quad \quad
\Delta U(z,z-\Delta z)=-qF(z) (1-\lambda) \Delta z,
\end{equation}
with $q$ the absolute value of the electron charge. For the sake of simplicity, $\lambda$ is assumed independent of $z$ and equal to $1/2$.

At a given position $z$, the carrier velocities in opposite directions around the energy $E_T$ are given by:
\begin{eqnarray}\label{v+-}
v_{\rightleftharpoons}(z,E_T)=\frac{\Delta z}{\tau_0}\exp\left[-\frac{E_C(z)-E_T}{kT_0}\right]\exp\left[\mp\frac{qF(z)\Delta z}{2kT_0}\right]
\end{eqnarray}
and, using Eq.\ (\ref{dnFermi}), their average values over the entire distribution are
\begin{equation}\label{avev}
\langle v_{\rightleftharpoons}(z)\rangle=\frac{1}{n(z)}\int_{E_V(z)}^{E_C(z)} \frac{v_{\rightleftharpoons}(z,E_T)}{1+\exp\left[\frac{E_T-E_F(z)}{kT(z)}\right]}\Gamma \,\dd E_T.
\end{equation}
The current density $J(z)$ thus reads:
\begin{equation}\label{defJ}
J(z)=-q\left[n\left(z-\frac{\Delta z}{2}\right)\left\langle v_{\rightarrow}\left(z-\frac{\Delta z}{2}\right)\right\rangle - n\left(z+\frac{\Delta z}{2}\right)\left\langle v_{\leftarrow}\left(z+\frac{\Delta z}{2}\right)\right\rangle\right].
\end{equation}
By expanding the r.h.s of Eq.\ (\ref{defJ}) to the first order in $\Delta z/2$, after some algebra we obtain:
\begin{equation}\label{Jgeneric}
\frac{J(z)}{q}=-n(z)\Big[\langle v_\rightarrow(z) \rangle - \langle v_\leftarrow(z) \rangle\Big]+\frac{\Delta z}{2}\frac{\dd}{\dd z} \left\{[n(z)\Big[\langle v_\rightarrow(z) \rangle + \langle v_\leftarrow(z) \rangle\Big]\right\},
\end{equation}
which, using Eqs.\ (\ref{dnFermi}), (\ref{v+-}) and (\ref{avev}), becomes:
\begin{eqnarray}\label{JFermi}
\frac{J(z)}{q}&=&-\frac{2\Delta z}{\tau_0}\sinh\left[-\frac{qF(z)\Delta z}{2kT_0}\right] A(z) + \frac{(\Delta z)^2}{\tau_0}\frac{\dd}{\dd z} \left\{ \cosh\left[-\frac{qF(z)\Delta z}{2kT_0}\right] \, A(z) \right\},
\end{eqnarray}
where 
$$A(z)=\int_{E_V(z)}^{E_C(z)} \exp\left[-\frac{E_C(z)-E_T}{kT_0}\right]\frac{\Gamma}{1+\exp\left[\frac{E_T-E_F(z)}{kT(z)}\right]}\dd E_T.
$$

If $J$ is fixed, as happens in the description of switching materials whose current-voltage characteristics are typically S-shaped, the model requires the determination of three unknown functions, namely the electric field $F(z)$, the quasi-Fermi level $E_F(z)$ and the carrier temperature $T(z)$. Two additional equations to be coupled to Eq.\ (\ref{JFermi}) are then required.

One of them is the Poisson equation, which, due to Eqs.\ (\ref{nFermi}) and (\ref{neq}), reads:
\begin{equation}\label{PoissonGeneric}
\frac{\dd F(z)}{\dd z}=\frac{\rho(z)}{\varepsilon}=-\frac{q}{\varepsilon}[n(z)-n_0]=\frac{q}{\varepsilon}\Gamma kT(z)\ln\left[\frac{1+\exp\frac{E_C(z)-E_F(z)}{kT(z)}}{1+\exp\frac{E_V(z)-E_F(z)}{kT(z)}}
\left(\frac{1+\exp\frac{E_{V0}-E_{F0}}{kT_0}}{1+\exp\frac{E_{C0}-E_{F0}}{kT_0}}\right)^{\frac{T_0}{T(z)}}
\right],
\end{equation}
where $\varepsilon=\varepsilon_0\varepsilon_r$ is the dielectric constant of the material. In writing Eq.\ (\ref{PoissonGeneric}) we have taken into account that the material under equilibrium conditions must be neutral everywhere. Since $E_{F0}$ has been assumed close to midgap and sufficiently far from the band edges, Eq.\ (\ref{PoissonGeneric}) simplifies into: 
$$
\frac{\dd F(z)}{\dd z}\approx\frac{q}{\varepsilon}\left\{\Gamma kT(z)\ln\left[\frac{1+\exp\frac{E_C(z)-E_F(z)}{kT(z)}}{1+\exp\frac{E_V(z)-E_F(z)}{kT(z)}}\right]-\frac{n_T}{2}\right\}.
$$

The third equation comes from the power balance. 
Depending on how effective electron-phonon scattering is in dissipating the power transferred to the carriers by the electric field, the average kinetic energy of the carriers (thus their temperature) may or may not stay tied to the equilibrium value. Let $\Delta E_{ex}^{TOT}(z)$ represent the \emph{excess energy}, i.e., the difference between the actual energy of the carrier distribution in $z$ and the energy that the same population would have if kept at the equilibrium temperature $T_0$:
\begin{equation}\label{DeltaEeccesso}
\Delta E_{ex}^{TOT}(z)=\int_{E_V(z)}^{E_C(z)}\frac{E_T-E_V(z)}{1+\exp\left[\frac{E_T-E_F(z)}{kT(z)}\right]}\Gamma \,\dd E_T - \int_{E_V(z)}^{E_C(z)}\frac{E_T-E_V(z)}{1+\exp\left[\frac{E_T-\tilde{E}_F(z)}{kT_0}\right]}\Gamma \, \dd E_T.
\end{equation}
Here $\tilde{E}_F(z)$ is defined at any $z$ by imposing the constraint
$$
\int_{E_V(z)}^{E_C(z)}\frac{1}{1+\exp\left[\frac{E_T-E_F(z)}{kT(z)}\right]}\dd E_T = \int_{E_V(z)}^{E_C(z)}\frac{1}{1+\exp\left[\frac{E_T-\tilde{E}_F(z)}{kT_0}\right]}\dd E_T,
$$
which ensures the same population for the two distributions.

In order to write the power balance, one considers the power flowing through two different sections of the device at a distance $\dd z$:
\begin{equation}\label{PowerGeneric}
 \Phi(z+\dd z)= \Phi(z)-J\dd \varphi-\left. \frac{\partial \Delta E_{ex}^{TOT}(z)}{\partial t}\dd z\right|_{\mbox{loss}}.
\end{equation}
Here $\Phi(z)$ is the energy density flux in $z$; $\dd\varphi=-F(z)\dd z$ is the variation of the electrostatic potential in the $z$ direction, and the last term of the r.h.s.\ represents the power exchanged via inelastic electron-phonon scattering. The derivative can be expressed in the relaxation-time approximation as\cite{JacoboniBook}
$$
\left.\frac{\partial{\Delta E_{ex}^{TOT}(z)}}{\partial t}\right|_{\mbox{loss}}=\frac{\Delta E_{ex}^{TOT}(z)}{\tau_r},
$$
$\tau_r$ being constant a relaxation time. After expanding the l.h.s.\ of Eq.\ (\ref{PowerGeneric}) to the first order in $\dd z$ and dividing both sides by $\dd z$, one gets
\begin{equation}\label{PowerBalance}
\frac{\dd \Phi(z)}{\dd z}=JF(z)-\frac{\Delta E_{ex}^{TOT}(z)}{\tau_r}.
\end{equation}

The energy density flux $\Phi(z)$ can be calculated following the same scheme adopted for the current density $J(z)$ in Eq.\ (\ref{defJ}):
\begin{equation}\label{defPhi}
\Phi(z)=n\left(z-\frac{\Delta z}{2}\right) \left \langle P_\rightarrow\left(z-\frac{\Delta z}{2}\right)\right \rangle - n\left(z+\frac{\Delta z}{2}\right) \left \langle P_\leftarrow\left(z+\frac{\Delta z}{2}\right)\right \rangle,
\end{equation}
where
\begin{equation}\label{aveP}
\langle P_{\rightleftharpoons}(z) \rangle=\frac{1}{n(z)}\int_{E_V(z)}^{E_C(z)}v_{\rightleftharpoons}(z,E_T) \frac{E_T-E_V(z)}{1+\exp\left[\frac{E_T-E_F(z)}{kT(z)}\right]}\Gamma \, \dd E_T
\end{equation}
represent the two average energy fluxes of the carrier distribution in opposite directions at a given coordinate $z$. 

The r.h.s.\ of Eq.\ (\ref{defPhi}) can be replaced with its first-order approximation in $\Delta z/2$, this leading to
$$
\Phi(z)=n(z)\Big [\langle P_{\rightarrow}(z)\rangle-\langle P_{\leftarrow}(z)\rangle\Big]-\frac{\Delta z}{2}\frac{\dd}{\dd z}\left\{n(z)\Big[\langle P_{\rightarrow}(z)\rangle+\langle P_{\leftarrow}(z)\rangle\Big]\right\}.
$$
Eq.\  (\ref{PowerBalance}) now reads:
\begin{equation}\label{PowerBalance2}
\frac{\dd}{\dd z}\left\{n(z)\Big[\langle P_{\rightarrow}(z)\rangle-\langle P_{\leftarrow}(z)\rangle\Big]\right\}-\frac{\Delta z}{2}\frac{\dd^2}{\dd z^2}\left\{n(z)\Big[\langle P_{\rightarrow}(z)\rangle+\langle P_{\leftarrow}(z)\rangle\Big]\right\}=JF(z)-\frac{\Delta E_{ex}^{TOT}(z)}{\tau_r},
\end{equation}
or, using Eqs. (\ref{dnFermi}), (\ref{v+-}), (\ref{DeltaEeccesso}) and (\ref{aveP}),

\begin{equation}\label{PhiFermi}
\begin{split}
&\frac{2\Delta z}{\tau_0} \frac{\dd}{\dd z} \left\{  \sinh\left[-\frac{qF(z)\Delta z}{2kT_0}\right] B(z)\right\} 
- \frac{(\Delta z)^2}{\tau_0}\frac{\dd^2}{\dd z^2} \left\{ \cosh\left[-\frac{qF(z)\Delta z}{2kT_0}\right] B(z)\right\}= \\
&\quad= JF(z)-\frac{1}{\tau_r} \left\{\int_{E_V(z)}^{E_C(z)}\Gamma \frac{E_T-E_V(z)}{1+\exp\left[\frac{E_T-E_F(z)}{kT(z)}\right]} \,\dd E_T - \int_{E_V(z)}^{E_C(z)}\Gamma \frac{E_T-E_V(z)}{1+\exp\left[\frac{E_T-\tilde{E}_F(z)}{kT_0}\right]} \, \dd E_T\right\},
\end{split}
\end{equation}
where
$$
B(z)=\int_{E_V(z)}^{E_C(z)} \Gamma \exp\left[-\frac{E_C(z)-E_T}{kT_0}\right] \frac{E_T-E_V(z)}{1+\exp\left[\frac{E_T-E_F(z)}{kT(z)}\right]}\,\dd E_T.
$$

The set of Eqs.\ (\ref{JFermi}), (\ref{PoissonGeneric}) and (\ref{PhiFermi}) leads to the determination of the unknown functions $F(z)$, $E_F(z)$ and $T(z)$ for any given current density $J$.
Eq.\ (\ref{PhiFermi}) involves the second derivatives of the unknown functions; nevertheless, a numerical analysis has shown that the term proportional to the second derivative in Eq.\ (\ref{PhiFermi}) is negligible with respect to the other term, but for a narrow region close to the injecting contact at the highest currents. For the latter case this term, though effective to some extent, still remains smaller than the other one in the l.h.s.\ of Eq.\ (\ref{PhiFermi}). The second-derivative contribution can thus be neglected without substantially affecting the physical results.

\begin{figure}
\includegraphics[width=.8\textwidth]{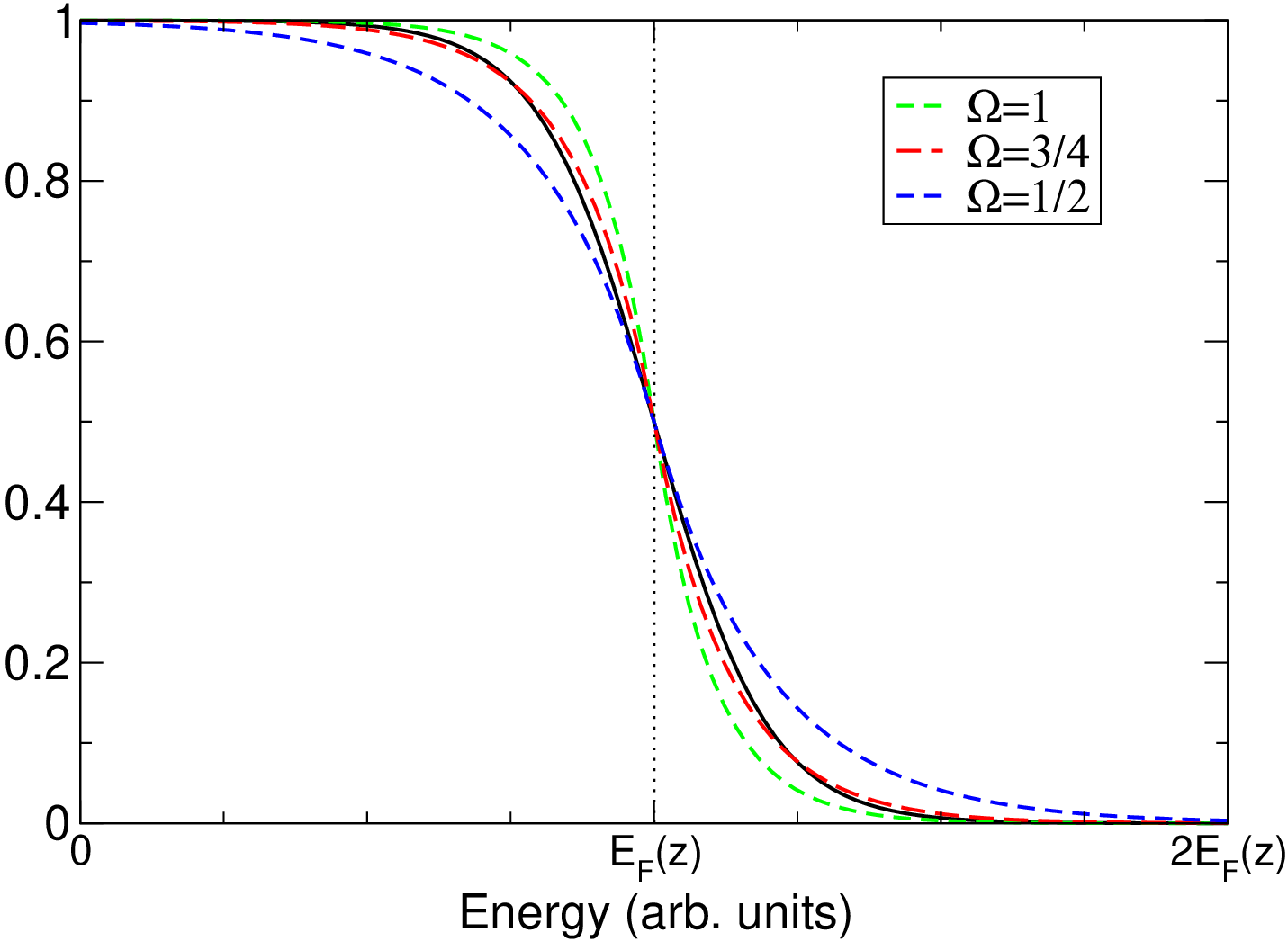}
\caption{(Color online) Comparison between the Fermi distribution function (solid dotted line) and the approximating function $\chi(E_T,z)$ for different values of the parameter $\Omega$ (dashed lines). The best approximation is obtained when $\Omega=3/4$.}
\label{distribuzioni}
\end{figure}

However, an analytical closed form for the solution of the set of Eqs.\ (\ref{JFermi}), (\ref{PoissonGeneric}) and (\ref{PhiFermi}) cannot be obtained because the integrals in Eq.\ (\ref{PhiFermi}) have to be evaluated numerically. One can overcome this problem by replacing the Fermi distribution function with a suitable approximation $\chi(E_T,z)$ defined in such a way that: i) $\chi(E_F(z))=1/2$, ii) $\chi(E_T,z)$ shares the same asymptotical values of the original Fermi-Dirac distribution, and iii) the following symmetry holds $1-\chi(E_F-\Delta)=\chi(E_F+\Delta)$. The above requirements are satisfied for instance by
\begin{equation}\label{chi}
\chi(E_T,z)= \left\{
\begin{array}{ll}
1-\frac{1}{2}\exp\left[\Omega\frac{E_T-E_F(z)}{kT(z)}\right]& \mbox{if }E_T< E_F(z)\\
&\\
\frac{1}{2}\exp\left[-\Omega\frac{E_T-E_F(z)}{kT(z)}\right]& \mbox{if }E_T\geq E_F(z)
\end{array}
\right.,
\end{equation}
where a good choice for the parameter $\Omega$ is $\Omega=3/4$, as shown in Fig.\ \ref{distribuzioni}.

Let us introduce the following dimensionless functions:
$$
f(z)=-\frac{qF(z)\Delta z}{2kT_0}, \quad \quad
g(z)=\frac{E_F(z)-E_{F0}}{kT_0}, \quad \quad
t(z)=\frac{T(z)}{T_0},
$$
which describe the off-equilibrium local field, the shift of the quasi-Fermi level, and the electron temperature, respectively.

After introducing the $\chi(E_T,z)$ function in place of the Fermi distribution, and neglecting the second derivative in Eq.\ (\ref{PhiFermi}), the set given by Eqs.\ (\ref{JFermi}), (\ref{PoissonGeneric}) and (\ref{PhiFermi}) can be manipulated through the straightforward, though lengthy, calculations summarized in the Appendix to yield:
\begin{align}
\label{ddf}
\frac{\dd f}{\dd z}&=N^*(g,t)\\
\label{ddg-ddt-1}
J_g(f,g,t)\frac{\dd g}{\dd z}+J_t(f,g,t)\frac{\dd t}{\dd z}&=J^*\left(f,g,t,\frac{\dd f}{\dd z}\right)\\
\label{ddg-ddt-2}
H_g(f,g,t)\frac{\dd g}{\dd z}+H_t(f,g,t)\frac{\dd t}{\dd z}&=H^*\left(f,g,t,\frac{\dd f}{\dd z}\right)
\end{align}
From the above, one immediately obtains:
\begin{equation}\label{Soluzione}
\frac{\dd f}{\dd z}=N^*, \quad \quad
\frac{\dd g}{\dd z}=\frac{J^*H_t-J_tH^*}{J_gH_t-J_tH_g}, \quad \quad
\frac{\dd t}{\dd z}=\frac{J_gH^*-J^*H_g}{J_gH_t-J_tH_g}.
\end{equation}
The definitions of the symbols can be found in the Appendix. 

Using a first-order Runge-Kutta integration scheme, the equations above are numerically solved once the values of $f(z), g(z)$ and $t(z)$ are provided at the coordinate $z=0$. Two boundary conditions can easily be inferred by supposing that the electrons at the injecting contact are at equilibrium, namely $E_F(0)=E_{F0}$ and $T(0)=T_0$. The definitions of $g(z)$ and $t(z)$ allow for the direct conversion of these boundary conditions into $g(0)=0$ and $t(0)=1$.

The value for $f(z)$ at the boundary must be such that the global charge neutrality of the system holds true:
\begin{equation}\label{vincoloQ}
\varepsilon\Big[F(0)-F(\ell)\Big]+q\int_0^{\ell} [n(z)-n_0]\,\dd z  =0.
\end{equation}
A trial-and-error procedure for the initial value $f(0)$ is applied, until the solutions for $f(z)$, $g(z)$ and $t(z)$ satisfy Eq.\ (\ref{vincoloQ}) within the desired precision.

\section{Results and Discussion}\label{Risultati}

\subsection{Current-voltage characteristics}

\begin{figure}
\includegraphics[width=.8\textwidth]{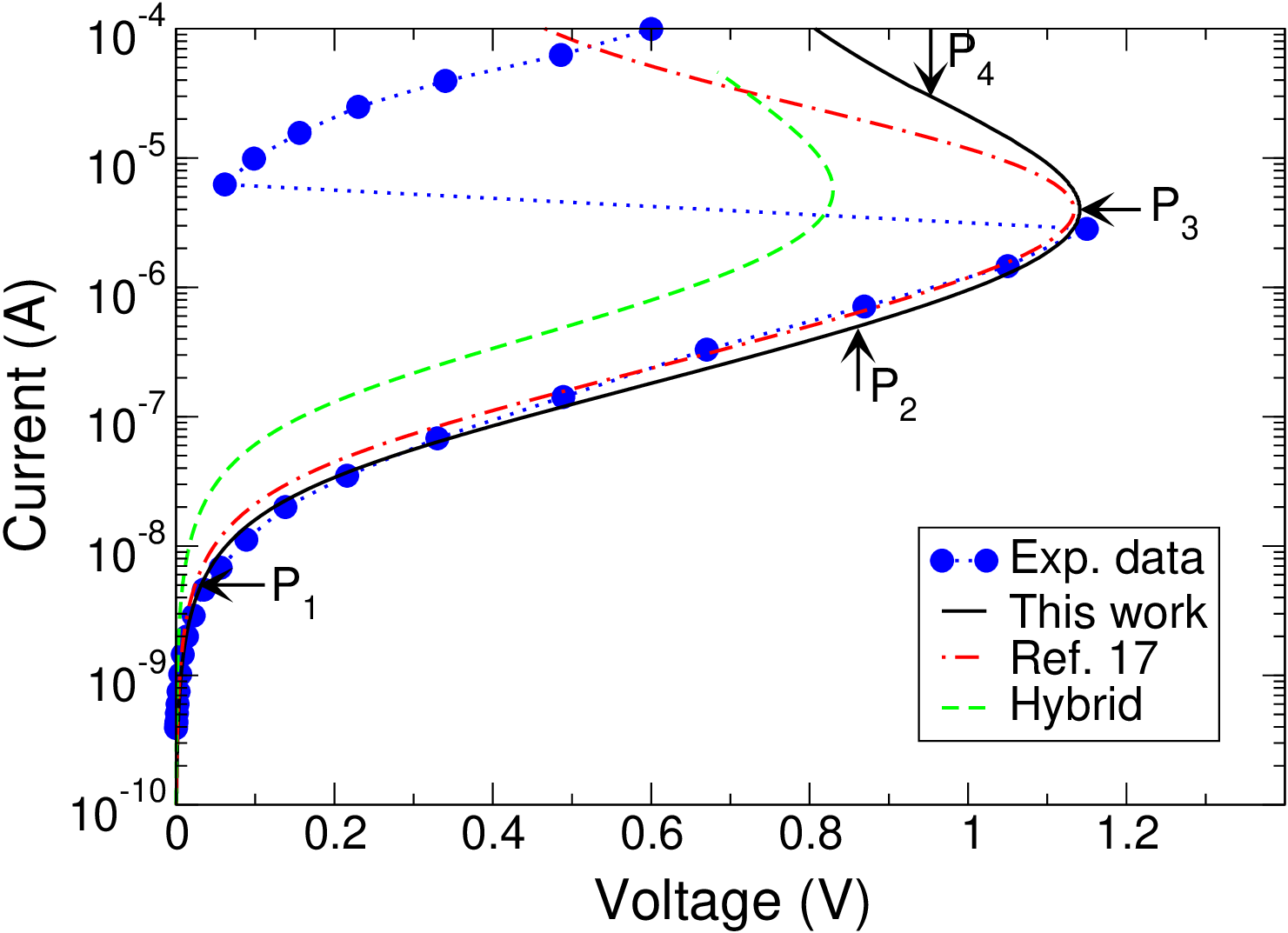}
\caption{(Color online) $I(V)$ characteristic obtained for the present model compared to experimental data and to the corresponding curve taken from Ref. \onlinecite{IelminiPRB}. The dashed green curve refers to a "hybrid" configuration where the parameters proposed in Ref. \onlinecite{IelminiPRB} have been used in the present model. The arrows show the positions along the $I(V)$ curve of the points P$_1$\dots P$_4$ cited in the text and in Figs.\ \ref{funzioniftn}.}
\label{caratteristiche}
\end{figure}

We report in Fig.\ \ref{caratteristiche} the $I(V)$ characteristic obtained with the model given by Eqs.\ (\ref{ddf}), (\ref{ddg-ddt-1}), and (\ref{ddg-ddt-2}) (solid black curve), where $V$ is calculated as the integral of the field $F(z)$ along the device. For comparison the result in Ref.\ \onlinecite{IelminiPRB} (dash-dotted red curve) and experimental data for a GST-225 memory cell available therein are also shown (dots). These data correspond to a memory cell with a bottom contact electrode with cross-section $\Sigma=1000$~nm$^2$ and length $\ell=40$~nm. The dashed green curve refers to a calculation performed using the present model with the same parameters reported in Ref.\ \onlinecite{IelminiPRB}.  It is clearly seen that, for the same set of parameters, the new model would lead to an evident increase in the conductivity (the dashed curve yields a given current at a lower voltage) and to a reduction of the threshold voltage, without substantially affecting the threshold current. These differences are due to the different approximation adopted for the carrier distribution function by the two models. While in Ref.\ \onlinecite{IelminiPRB} the carriers that contribute to transport are only those above the Fermi level, in the present model the Fermi distribution function is integrated over the entire band gap, this making the conductivity higher. Since the threshold point is determined mainly by the current, as will be discussed later on, the increased conductivity implies a smaller threshold voltage.

In order to fit the experimental data, it has then been necessary to calibrate the parameters. The new parameter set is reported in Table \ref{parameterlist} along with the set used in Ref.\ \onlinecite{IelminiPRB} for the sake of comparison.
The band gap considered here is compatible with literature data for the amorphous GST-225. The two time constants $\tau_0$ and $\tau_r$ are consistent with those suggested by Mott and Davis for amorphous semiconductors.\cite{MottBook}  
\begin{table}[htbp]
\begin{tabular}{l|c|c}
parameter&present work&Ref.\ \onlinecite{IelminiPRB}\\
\hline\hline
$\Delta E_G$ &0.68 eV & 0.6 eV\\
\hline
$n_T/\Delta E_G$ & $10^{20}$ cm$^{-3}$eV$^{-1}$& $10^{20}$ cm$^{-3}$eV$^{-1}$\\
\hline
$\tau_0$&$1.2\cdot10^{-14}$ s &$1.0\cdot10^{-14}$ s\\
\hline
$\tau_r$&$0.78\cdot10^{-13}$ s &$1.0\cdot10^{-13}$ s\\
\hline
$\Delta z$& $7\cdot10^{-7}$ cm& $7\cdot10^{-7}$ cm\\
\hline
$\varepsilon_r$& 15 &15\\
\hline
$T_0$ & 298 K & 298 K
\end{tabular}
\caption{Parameters used for the best fit reported in Fig.\ \ref{caratteristiche}. The values used in Ref.\ \onlinecite{IelminiPRB} are also reported for comparison.}
\label{parameterlist}
\end{table}

The agreement between the results of the present model with the new set of parameters and the experimental data of is quite good. Differences with the results of model of Ref.\ \onlinecite{IelminiPRB} are found in the NDR region, where the present model estimates larger potential drops for any given current. We point out, however, that above the switching point the experimental data depend on the characteristics of the external circuit. For this reason they should not be considered as for the sake of comparison. In fact, under different experimental conditions the current may rise nearly vertically at a holding voltage.\cite{Fritzsche} Then, the crystallization of the material occurs. Results from the present model obtained far above threshold should thus be considered only qualitatively.

\subsection{Microscopic interpretation of the switching}

\begin{figure}[htbp]
\centering
\includegraphics*[width=0.8\textwidth]{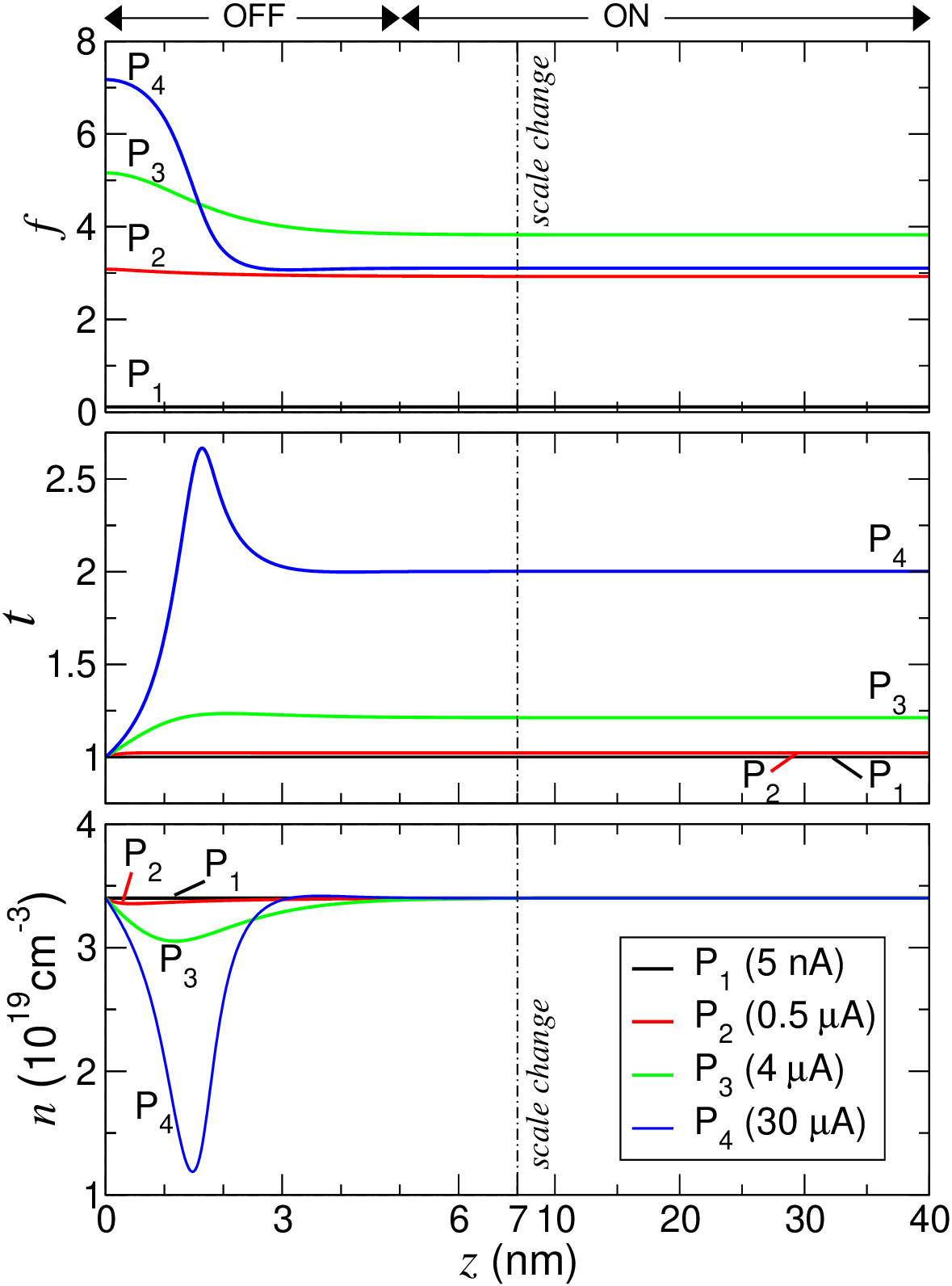}
\caption{(Color online) Dimensionless electric field $f(z)$ (top), dimensionless electron temperature $t(z)$ (middle), and electron concentration $n(z)$ (bottom) for the four currents identified in Fig.\ \ref{caratteristiche}. Note the scale change between the OFF and the ON regions at $z=7$~nm.} 
\label{funzioniftn}
\end{figure}

In order to analyze the microscopic process leading to the threshold switching, it is useful to consider the physical quantities of interest along the device. Apart from the amplitude of the variations, a common behavior for the dimensionless electric field $f(z)$ and carrier temperature $t(z)$, and for the carrier concentration $n(z)$, can be outlined. The calculated profiles for these quantities are reported in Fig.\ \ref{funzioniftn} for the four points P$_1$\dots P$_4$ shown in Fig.\ \ref{caratteristiche}.\cite{SuppMatMovie}

Since the injecting contact acts as an infinite reservoir of carriers in thermal equilibrium with the lattice, carriers enter the device at thermal equilibrium. However, due to the imbalance between the power provided by the field and the power loss due to electron-phonon scattering, they tend to heat up as long as they travel along the device. This effect is negligibly small at the lowest currents. The increase in the carrier temperature would enhance the flux towards the collecting contact and, in order to keep $J$ at its prescribed value, it is compensated by a decrease in the carrier concentration and in the electric field, consistently with the Poisson equation. Thus, a positive charge accumulates near the injecting contact. The position of the quasi-Fermi level shifts towards a lower energy to account for such a change in the carrier concentration.
 
Going farther from the injecting contact, the heated carrier population enables a more effective dissipation through inelastic scattering, which prevails over the power provided by the electric field. The electron temperature, after reaching a maximum value, decreases. A thermal overshoot is thus created near the injecting contact. Since the charge must vary continuously in space, the presence of a positive charge implies that the electric field continues to decrease also in the region where the carrier temperature is reduced, although at a lower rate. In order to preserve the current, the carrier concentration must compensate the reduction in velocity due to the smaller field and temperature, so that it increases significantly and approaches again the equilibrium value. When the charge neutrality is attained far enough from the injecting contact, the electric field and the carrier temperature saturate as well. It follows that a saturation value for the position of the quasi-Fermi level is also found. 

As in Ref.\ \onlinecite{IelminiJAP}, we refer to the OFF region as the zone of length $\ell_{OFF}$ close to the injecting contact where the physical quantities vary most appreciably, and to the ON region as the longer zone of length $\ell_{ON}$ where they have their saturation values, as shown in Fig.\ \ref{funzioniftn}.

Even though the interpretation given above applies at any current, in the first part of the characteristic (point P$_1$, $I=5$~nA) the electric field is low and does not provide enough power to determine an appreciable electron heating. This fact implies that the electron temperature and concentration are tied to their equilibrium values across the whole device (Ohmic behavior).

A similar situation applies also for the most part of the subsequent exponential region. However, as the current increases, the electric field $f(0)$ grows rapidly (P$_2$, $I=0.5$~$\mu$A) and eventually triggers an appreciable electron heating. In turn, the latter causes a slight depletion of the OFF region and a corresponding weak non-uniformity of the electric field. Due to the exponential relationship linking the current and the field, this picture is more and more evident as the switching current is approached  (P$_3$, $I=4$~$\mu$A).
 
As the current is increased above the threshold point (P$_4$, $I=30$~$\mu$A), $f(0)$ still continues to grow, thus inducing a larger carrier heating in the OFF region. As a consequence the local electric field rapidly falls, and the electron concentration is strongly depleted to keep the current constant. The high value of the electron temperature in the OFF region reflects into a high saturation value also in the ON region, which is the key condition to restore the equilibrium between energy gain and energy relaxation. Since in the ON region the carrier concentration has always the equilibrium value at any current, every further increase in the carrier temperature can only be compensated by a further decrease in the electric field. For such a reason, the resulting electric field is smaller than that found at the switching point. Since the ON region is substantially much longer than the OFF region, a smaller field in the ON region leads also to smaller potential drop across the device. 

Finally, in the region above the switching point the power density dissipated and transferred to the lattice via electron-phonon scattering is high. This may give rise to lattice heating, which is a favorable condition for the creation of a local crystalline nucleus that can eventually evolve into a crystalline filament. The incorporation of the Fourier heat equation and a local lattice temperature in the model goes beyond the scope of the present paper and is planned for a future work.  

In conclusion, few words must be spent on the role of the Poisson equation, by comparing the present results with those of Ref.\ \onlinecite{IelminiPRB}. As in the ON region charge neutrality is kept, both models provide similar results; on the contrary, in the OFF region the introduction of a self-consistent non-uniform carrier distribution yields a more accurate physical picture. In particular, the presence of a minimum in the carrier concentration profile implies a thermal overshoot and an initial different curvature of the field profile.

\subsection{The switching condition}

\begin{figure}[htbp]
\includegraphics*[width=.8\textwidth]{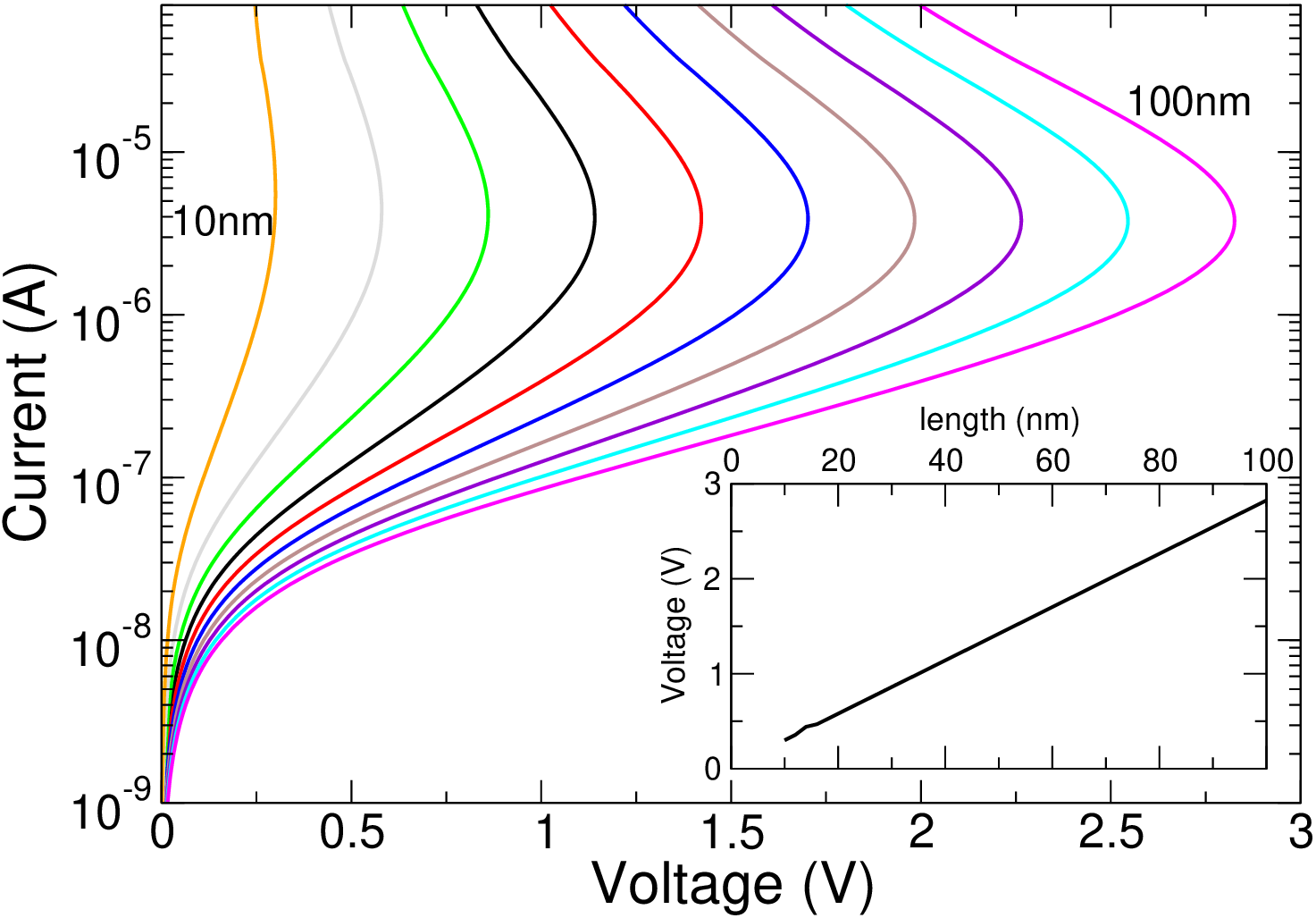}
\caption{(Color online) $I(V)$ characteristics for devices with different lengths ranging from 10~nm to 100~nm with steps of 10~nm. The inset shows the linear relationship between the threshold voltage and the device length.}
\label{CaratteristicheL}
\end{figure}

\begin{figure}[htbp]
\includegraphics*[width=.8\textwidth]{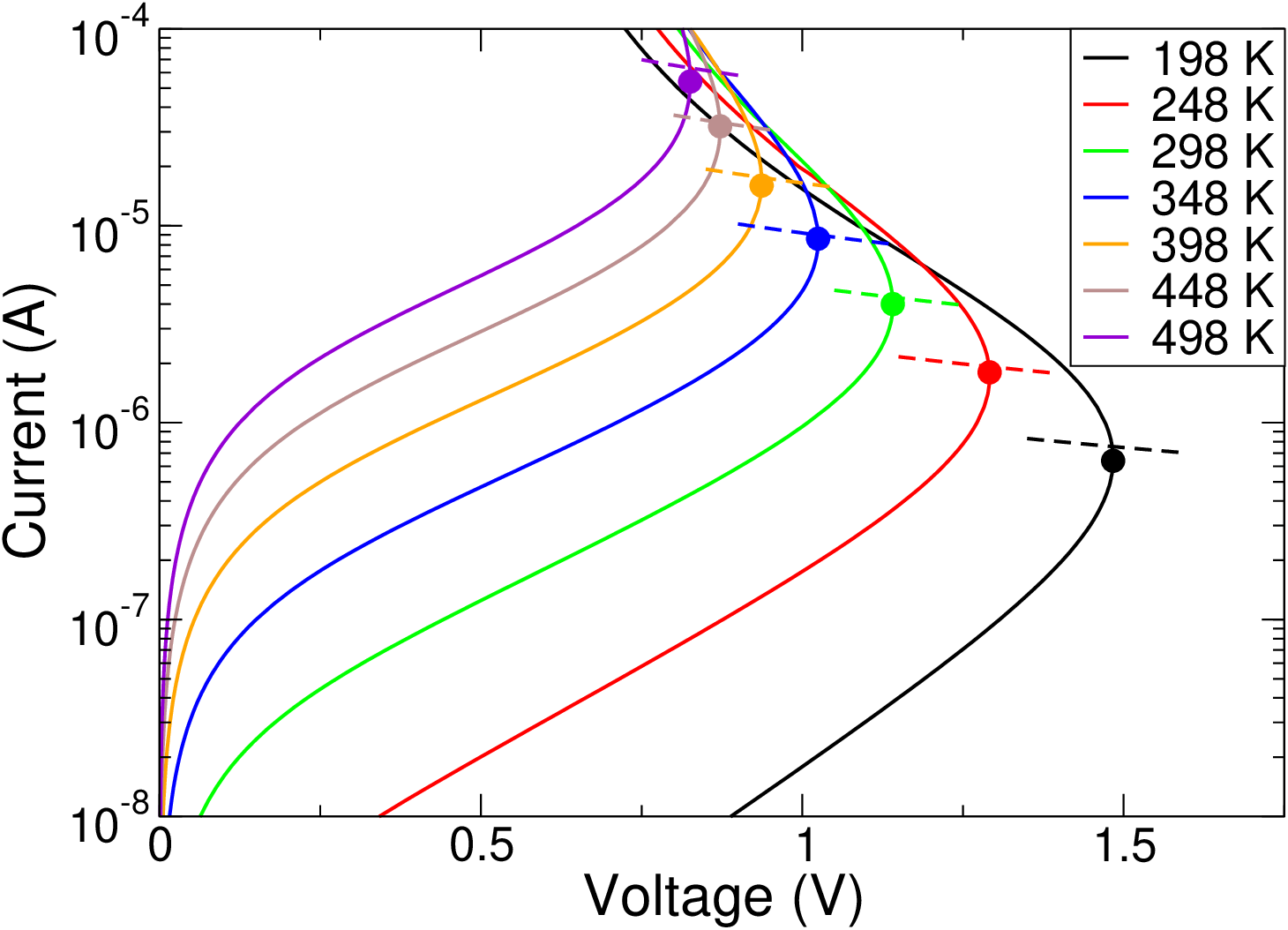}
\caption{(Color online) $I(V)$ characteristics for different lattice temperatures in the range 198~K to 498~K.  The intersections with the dashed lines represent the switching points as calculated from Eq.\ (\ref{Wthapprox}).}
\label{CaratteristicheT}
\end{figure}

The last part of this section is devoted to the analysis of the switching condition as a function of the device length (Fig.\ \ref{CaratteristicheL}) and the lattice temperature (Fig.\ \ref{CaratteristicheT}). 

The OFF region, as shown in Fig.\ \ref{funzioniftn}, extends over about the first 5~nm from the injecting contact, which is the space where carriers cannot fully relax the power provided by the field through electron-phonon scattering and heat up. The microscopic phenomena occurring in the OFF region suggest that $\ell_{OFF}$ must primarily depend on material properties, like the density of traps $n_T$, the position of the Fermi level with respect to the bottom of the conduction band, and the relaxation time $\tau_r$, but must be independent of the device length. 
As a consequence, when the latter exceeds approximately $2\ell_{OFF}$ the potential drop in the ON region dominates over that in the OFF region, and the threshold voltage scales almost linearly with the device length, as shown in the inset of Fig.\ \ref{CaratteristicheL}. In this case, the error made by considering the electric field $F_{th}$ in the ON region as representative of the field in the entire device at threshold is negligible. On the other hand, small deviations from linearity are found for shorter devices, as the potential drop in the OFF region gains relative importance over that in the ON region. According to the present model, ultra-short devices are not expected to show a NDR portion of the $I(V)$ characteristic, but, rather, a steep rise of the current with an almost constant potential, in agreement with the results of Fig.\ \ref{CaratteristicheL}.

In order to assess the effect of the lattice temperature on the switching condition, we report in Fig.\ \ref{CaratteristicheT} the $I(V)$ characteristics for different lattice temperatures. It is found that the threshold current increases with temperature, while the threshold potential decreases, as typical of chalcogenide glasses.\cite{FritzscheIBM} Moreover, the calculations show that the threshold potentials tend to accumulate towards a limiting value as the temperature increases, suggesting the existence of a minimum threshold electric field.

With the purpose of understanding this phenomenon, we recall first that the ON region is defined as the zone where the equilibrium carrier concentration is restored, and the electric field and carrier temperature saturate. This physical condition is expressed by making the l.h.s.\ of Eq.\ (\ref{PowerBalance2}) to vanish; viz.,
$$
JF(z)-\frac{\Delta E_{ex}^{TOT}(z)}{\tau_r}=0.
$$

\begin{figure}[htbp]
\includegraphics*[width=.8\textwidth]{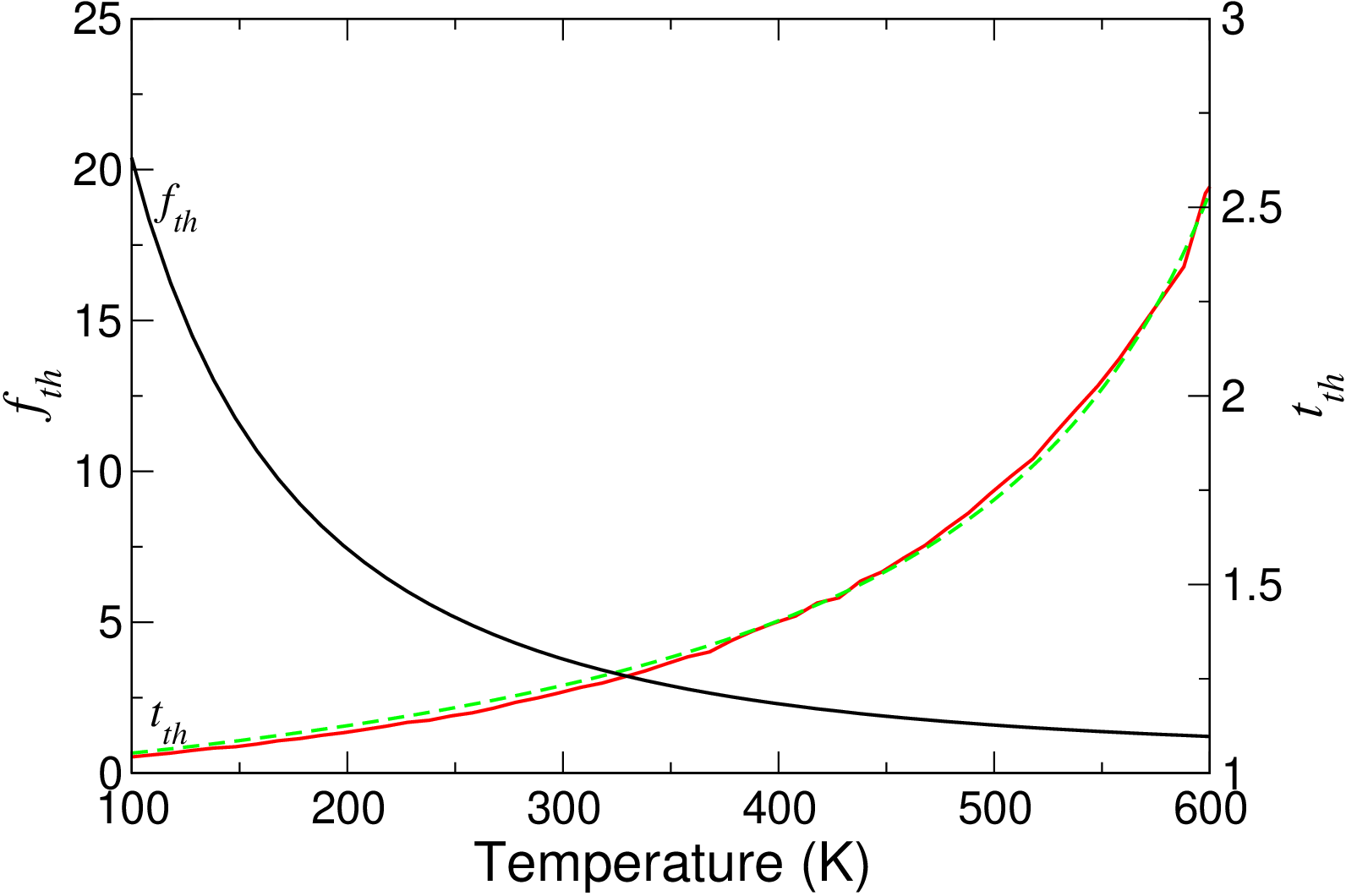}
\caption{(Color online) Values at the threshold point of the dimensionless electric field and of the carrier temperature. The dashed green line is the best-fit approximation of $t_{th}$ obtained by means of Eq.\ (\ref{tthapprox}).}
\label{FunzioniSogliaT}
\end{figure}

Taking $F_{th}$ as representative of the field within the device, the product $J_{th}F_{th}$ is the input power density $w_{th}$ at the threshold point. Here and in the following, the suffix $th$ indicates the quantities evaluated at the threshold point. The definitions of $\Delta E_{ex}^{TOT}(z)$ and $\U$ given by Eqs.\ (\ref{Eapprox}) and (\ref{U}) in the Appendix yield:
$$
w_{th}=\frac{I_{th}}{\Sigma}\frac{\Delta\varphi_{th}}{\ell}=\Gamma\,(kT_0)^2\,\frac{t_{th}^2 \U\Big|_{th}-U_0(\alpha_0^{\prime},\beta_0^{\prime})\Big|_{th}}{\tau_r}.
$$
If $t_{th}<2.5$, as happens for a wide range of lattice temperatures up to 600~K (see also Fig.\ \ref{FunzioniSogliaT}), the exponents in $\U$ can be neglected, and the above equation further simplifies to
\begin{equation}\label{ThPower}
w_{th}=\frac{16}{9}\,\Gamma k^2\,\frac{T_{th}^2-T_0^2}{\tau_r},
\end{equation}
where the condition that the quasi-Fermi level in the ON region is close to the local equilibrium value ($g_{th}\approx 0$) has also been used. The lack of an analytical solution for $t(z)$ makes it impossible to further simplify Eq.\ (\ref{ThPower}) into a compact form showing the dependences  on the parameters of the material only. However, by means of numerical analyses it is possible to obtain an empirical formula linking $t_{th}$ with the lattice temperature. 

Since the threshold current strongly increases with temperature overcompensating the decrease in the threshold potential, the input power at threshold increases indefinitely. The above considerations suggest an approximate function for $t_{th}$ like, e.g.,
\begin{equation}\label{tthapprox}
t_{th}=\frac{\theta T_0 - T^*}{T_0 - T^*},
\end{equation}
where the parameter $T^*$ has the physical meaning of an asymptotical temperature up to which the NDR regions of the $I(V)$ characteristics are possible. For the case at hand, by best fitting the data reported in Fig.\ \ref{FunzioniSogliaT} (dashed green line), one finds $\theta\approx 2/3$ and $T^*\approx 730$~K, a value slightly larger than the glass transition temperature of the material. 
We point out, however, that these values should be considered with care, as they depend non-linearly on the parameters of the model. Numerical calculations have shown that a major role is played by the energy relaxation time $\tau_r$, which significantly influences the threshold current and voltage. In fact, increasing the energy relaxation time by one order-of-magnitude reduces the asymptotical temperature by about 100~K, whereas reducing the relaxation time by one order-of-magnitude let the asymptotical temperature raise only by 35~K. The $\theta$ coefficient ranges instead from 0.75 down to 0.63 for the same variations of $\tau_r$.

By inserting Eq.\ (\ref{tthapprox}) into Eq.\ (\ref{ThPower}), one finds:
\begin{equation}\label{Wthapprox}
w_{th}=\frac{16\, n_T\, (kT_0)^3}{9\, \tau_r\,\Delta E_G}\frac{(\theta+1)kT_0-2kT^*}{(kT_0-kT^*)^2}(\theta-1)
\end{equation}
that represents the critical power at threshold as a function of the lattice temperature. In order to provide an estimate of the switching point, this formula must be combined with Eq.\ (\ref{jdriftapprox}) that expresses the current in the ON region. Taking again into account that the quasi-Fermi level in the ON region is close to its local equilibrium value, and using Eq.\ (\ref{tthapprox}) to express the carrier temperature at threshold, it is possible to obtain the threshold current as a function uniquely of the dimensionless field $f_{th}$. By inserting the expression for $J_{th}$ from Eq.\ (\ref{jdriftapprox}) into Eq.\ (\ref{Wthapprox}), after some algebra, one gets:
\begin{equation}\label{fth}
f_{th}\sinh(f_{th})=\frac{8}{9}\frac{\tau_0}{\tau_r}\exp\left(\frac{\Delta E_G}{2kT_0}\right)kT_0\frac{(\theta+1)kT_0-2kT^*}{(kT_0-kT^*)^2}\frac{\theta-1}{Q_{th}(T_0,T^*,\theta)},
\end{equation}
where $Q_{th}(T_0,T^*,\theta)$ is calculated by setting $t=t_{th}$ and $g=0$ in Eq.\ (\ref{Q}). Once $f_{th}$ is known from Eq.\ (\ref{fth}), the threshold potential and current are given as $\Delta\varphi_{th}=F_{th}\ell$ and $J_{th}=w_{th}/F_{th}$, respectively. The above results allow for a rough estimate of the threshold point from the physical parameters of a given switching material.

\section{Conclusions}

An enhanced model for transport in amorphous chalcogenides has been worked out to achieve self-consistency between the electric field present along the device and the local density and energy distribution of carriers. This characteristic was, in fact, missing in the models available in the literature.

The model relies on the hypothesis that conduction can be described by means of trap-limited transport, i.e., sequences of detrapping events letting a carrier move above the conduction-band mobility gap, followed by fast recaptures by different traps. 
Within this framework, it is possible to write two equations for the charge and the energy fluxes where two characteristic times are present: the detrapping time and the energy relaxation time. A third equation expresses charge-field consistency.
The carrier concentration and the charge and energy fluxes are obtained by means of the integration of the Fermi distribution function over the band gap, which is the energy region filled by trap states. The solution of the above set of equations yields the three main physical quantities that represent the unknowns of the problem, namely, the carrier concentration, the carrier temperature, and the electric field as functions of the position along the device.

After a proper identification of the physical parameters, the calculated current-voltage characteristics are in good agreement with experimental data. Moreover, the model correctly identifies both the temperature and device-length dependences of the $I(V)$ curves.

The ovonic switching behavior of chalcogenide glasses is explained by the effect of carrier heating. 
In trap-limited conduction, conductivity is determined mainly by the detrapping time, which depends on the energy barrier separating the energy of the carrier sitting in a trap from the conduction-band mobility edge. Detrapping processes are therefore favored by a lowering of the barrier due to the electric field and by the raising of the carrier energy due to their heating. At and above threshold, any further increase in the current requires a higher carrier temperature, which reduces the field in the largest part of the device, so that the potential drop decreases. The model can thus predict the switching point starting from the physical parameters of the material, which is a key achievement for technological applications.
 
\begin{acknowledgments}
Part of this work has been carried out under the contract 3477131/2011 of the Intel Corp.\ whose support is gratefully acknowledged.
\end{acknowledgments}
 
\appendix*
\section{Analytical derivation of the constitutive equations}\label{AppendiceEquazioni}

This appendix contains in some details the calculations which transform Eqs. (\ref{JFermi}), (\ref{PoissonGeneric}) and (\ref{PhiFermi}) in the main text into Eqs. (\ref{ddf}), (\ref{ddg-ddt-1}) and (\ref{ddg-ddt-2}) so that the interested reader can follow the mathematical development between the two sets of equations.

Let us introduce the dimensionless auxiliary variables:
$$
\alpha_0(z)=C_0+g(z), \quad \quad
\alpha(z)=\frac{\alpha_0}{t(z)}, \quad \quad 
\beta_0(z)=C_0-g(z), \quad \quad
\mbox{and} \quad \quad
\beta(z)=\frac{\beta_0}{t(z)},$$
with $C_0=\Delta E_G/2kT_0$. Following the calculations sketched in Sect.\ \ref{model}, after replacing the Fermi-Dirac distribution with $\chi(E_T,z)$, one finds (omitting the indication of the dependence on $z$ of the unknown functions):
\begin{eqnarray}
\label{napprox}
n&=&\int_{E_V}^{E_C}\Gamma \chi(E_T,z)\, \dd E_T=\Gamma kT_0\, t \N\\
\label{jdriftapprox}
n\Big(\langle v_\leftarrow\rangle - \langle v_\rightarrow\rangle\Big)&=&\int_{E_V}^{E_C}\Gamma \chi(E_T,z)\frac{\Delta z}{\tau_0}\exp\left(-\frac{E_C-E_T}{kT_0}\right)\sinh\left(-\frac{qF\Delta z}{2kT_0}\right)\, \dd E_T=\nonumber\\
&=&\frac{\Gamma kT_0 \Delta z}{\tau_0}\, \sinh(f)\exp(-\beta_0)\Q\\
\label{jdiffapprox}
n\Big(\langle v_\leftarrow\rangle + \langle v_\rightarrow\rangle\Big)&=&\int_{E_V}^{E_C}\Gamma \chi(E_T,z)\frac{\Delta z}{\tau_0}\exp\left(-\frac{E_C-E_T}{kT_0}\right)\cosh\left(-\frac{qF\Delta z}{2kT_0}\right)\, \dd E_T=\nonumber\\
&=&\frac{\Gamma kT_0 \Delta z}{\tau_0}\, \cosh(f)\exp(-\beta_0)\Q\\
\label{Eapprox}
\Delta E_{ex}^{TOT}&=&\int_{E_V}^{E_C}\Gamma \Big[\chi(E_T,z)-\tilde{\chi}(E_T,z)\Big|_{T=T_0}\Big](E_T-E_V)\, \dd E_T=\nonumber\\
&=&\Gamma (kT_0)^2\, \Big[t^2\U-U_0(\alpha_0^{\prime},\beta_0^{\prime})\Big]\\
\label{Pdriftapprox}
n\Big(\langle P_\leftarrow\rangle - \langle P_\rightarrow\rangle\Big)&=&\int_{E_V}^{E_C}\Gamma \chi(E_T,z)(E_T-E_V)\frac{\Delta z}{\tau_0}\exp\left(-\frac{E_C-E_T}{kT_0}\right)\sinh\left(-\frac{qF\Delta z}{2kT_0}\right)\, \dd E_T=\nonumber\\
&=&\frac{2\Gamma\Delta z (kT_0)^2}{\tau_0}\, \sinh(f)\exp(-\beta_0)\S\\
\label{Pdiffapprox}
n\Big(\langle P_\leftarrow\rangle + \langle P_\rightarrow\rangle\Big)&=&\int_{E_V}^{E_C}\Gamma \chi(E_T,z)(E_T-E_V)\frac{\Delta z}{\tau_0}\exp\left(-\frac{E_C-E_T}{kT_0}\right)\cosh\left(-\frac{qF\Delta z}{2kT_0}\right)\, \dd E_T=\nonumber\\
&=&\frac{2\Gamma\Delta z (kT_0)^2}{\tau_0}\, \cosh(f)\exp(-\beta_0)\S
\end{eqnarray}
with
\begin{equation}\label{N}
\N=\alpha+\frac{2}{3}\exp\left(-\frac{3}{4}\alpha\right)-\frac{2}{3}\exp\left(-\frac{3}{4}\beta\right),
\end{equation}
\begin{equation}\label{Q}
\Q=\frac{18}{9-16t^2}-\exp(-\alpha_0)\left[2-\frac{4t}{4t+3}\exp\left(-\frac{3}{4}\alpha\right)\right]+\frac{4t}{4t-3}\exp\left(\beta_0-\frac{3}{4}\beta\right),
\end{equation}
\begin{equation}\label{U}
\U=\frac{16}{9}+\frac{\alpha^2}{2}-\frac{8}{9}\exp\left(-\frac{3}{4}\alpha\right)-\frac{4}{3}\left(\frac{C_0}{t}+\frac{2}{3}\right)\exp\left(-\frac{3}{4}\beta\right),
\end{equation}
and
\begin{eqnarray}\label{S}
\S&=&\alpha_0\frac{9}{9-16t^2}+9\frac{48t^2-9}{(9-16t^2)^2}+\exp(-\alpha_0)\left[1-\frac{1}{2}\left(\frac{4t}{4t+3}\right)^2\exp\left(-\frac{3}{4}\alpha\right)\right]+\nonumber \\
&&+\frac{4t}{4t-3}\left(C_0-\frac{2t}{4t-3}\right)\exp\left(\beta_0-\frac{3}{4}\beta\right).
\end{eqnarray}

The $U_0(\alpha_0^{\prime},\beta_0^{\prime})$ function in Eq.\ (\ref{Eapprox}) can be calculated by setting $(t,\alpha,\beta)=(1,\alpha_0^{\prime},\beta_0^{\prime})$ in $\U$, with $\alpha_0^{\prime}$ and $\beta_0^{\prime}$ fulfilling the constraints $ t \N=N(\alpha_0^{\prime},\beta_0^{\prime})$ and $\alpha_0^{\prime}+\beta_0^{\prime}=2C_0$.

Under equilibrium it is $f=0$, $g=0$, and $t=1$. As a consequence, $\alpha_{eq}=\beta_{eq}=C_0$. The equilibrium electron concentration thus reads:
$$
n_{0}=\Gamma kT_0\,C_0=\frac{n_T}{2}.
$$
By means of this relationship and of Eq.\ (\ref{napprox}), Eq.\ (\ref{PoissonGeneric}) can then be recast in terms of the dimensionless auxiliary variables as follows:
\begin{equation}\label{PoissonApprox}
\frac{\dd f}{\dd z}=C_1\Big[t \N-C_0\Big]
\end{equation}
with $C_1=q^2\Gamma \Delta z/(4\varepsilon)$.

Eq.\ (\ref{JFermi}) can be approximated by means of Eqs.\ (\ref{Jgeneric}), (\ref{jdriftapprox}) and (\ref{jdiffapprox}). Recalling that
$$
\frac{\dd \alpha_0}{\dd z}=\frac{\dd g}{\dd z}, \quad  
\frac{\dd \alpha}{\dd z}=\frac{1}{t}\left(\frac{\dd g}{\dd z}-\alpha\frac{\dd t}{\dd z}\right), \quad 
\frac{\dd \beta_0}{\dd z}=-\frac{\dd g}{\dd z}, \quad 
\mbox{and} \quad 
\frac{\dd \beta}{\dd z}=-\frac{1}{t}\left(\frac{\dd g}{\dd z}+\beta\frac{\dd t}{\dd z}\right),
$$
and observing also that
$$
\frac{\partial Q}{\partial \alpha_0}=2\exp(-\alpha_0)+\frac{4}{3}\frac{\partial Q}{\partial \alpha}, \quad \quad
\mbox{and } \quad \quad
\frac{\partial Q}{\partial \beta_0}=-\frac{4}{3}\frac{\partial Q}{\partial \beta},
$$
Eq.\ (\ref{JFermi}) can be recast as
\begin{equation}\label{Japprox}
J_g\frac{\dd g}{\dd z}+J_t\frac{\dd t}{\dd z}=\sinh(f)\Q\left(\frac{2}{\Delta z}-\frac{\dd f}{\dd z}\right)+C_2\,J\exp(\beta_0),
\end{equation}
where
$$
J_g=\cosh(f)\left[\Q+2\exp(-\alpha_0)+\left(\frac{4}{3}+\frac{1}{t}\right)\frac{\partial Q}{\partial \alpha}+\left(\frac{4}{3}-\frac{1}{t}\right)\frac{\partial Q}{\partial \beta}\right],
$$
$$
J_t=\cosh(f)\left(\frac{\partial Q}{\partial t}-\frac{\alpha}{t}\frac{\partial Q}{\partial \alpha}-\frac{\beta}{t}\frac{\partial Q}{\partial \beta}\right),
$$
and $C_2= 2\tau_0 /[q \Gamma kT_0 (\Delta z)^2]$.

Similarly, after neglecting the second derivative as indicated in the text, Eq.\ (\ref{PhiFermi}) can be recast making use of Eqs.\ (\ref{PowerBalance2}) and (\ref{Pdriftapprox}), this yielding: 
\begin{equation}\label{PowerBalanceApprox}
H_g\frac{\dd g}{\dd z}+H_t\frac{\dd t}{\dd z}=-\exp(\beta_0)\Big\{2C_2Jf-C_3\Big[t^2\U-U_0(\alpha^{\prime},\beta^{\prime})\Big]\Big\}-H_f\frac{\dd f}{\dd z},
\end{equation} 
with 
$$
H_f=\cosh(f)\S,
$$
$$
H_g=\sinh(f)\left[\S+\frac{1}{t}\left(\frac{\partial S}{\partial \alpha}-\frac{\partial S}{\partial \beta}\right)+\frac{\partial S}{\partial \alpha_0}-\frac{\partial S}{\partial \beta_0}\right],
$$
$$
H_t=\sinh(f)\left(\frac{\partial S}{\partial t}-\frac{\alpha}{t}\frac{\partial S}{\partial \alpha}-\frac{\beta}{t}\frac{\partial S}{\partial \beta}\right),
$$
where $C_3=\tau_0/(4\Delta z \tau_r)$. 

Eqs.\ (\ref{PoissonApprox}), (\ref{Japprox}) and (\ref{PowerBalanceApprox}) can be turned into Eqs.\ (\ref{ddf}), (\ref{ddg-ddt-1}) and (\ref{ddg-ddt-2}) by letting
$$
N^*=C_1\Big[t \N-C_0\Big],
$$
$$
J^*=\sinh(f)\Q\left(\frac{2}{\Delta z}-\frac{\dd f}{\dd z}\right)+C_2\,J\exp(\beta_0),
$$
and
$$
H^*=\exp(\beta_0)\Big\{2C_2Jf-C_3\Big[t^2\U-U_0(\alpha^{\prime},\beta^{\prime})\Big]\Big\}-H_f\frac{\dd f}{\dd z}.
$$

\end{document}